\documentclass{article}

\PassOptionsToPackage{round}{natbib}

\usepackage[final]{neurips_2018}

\usepackage[utf8]{inputenc} 
\usepackage[T1]{fontenc}    
\usepackage{hyperref}       
\usepackage{url}            
\usepackage{booktabs}       
\usepackage{amsfonts}       
\usepackage{nicefrac}       
\usepackage{microtype}      

\usepackage{graphicx,colortbl}
\usepackage{float}
\usepackage{amsmath}
\usepackage{amssymb}
\usepackage{amsthm}
\usepackage{calc}
\usepackage{mathtools}
\usepackage{bbm}
\usepackage{cleveref}
\usepackage{caption} 
\usepackage{subcaption}
\usepackage{multirow}
\usepackage[linesnumbered,ruled,vlined]{algorithm2e}

\usepackage{environ}
  {\renewcommand\section[1]{\refstepcounter{section}}%
   \renewcommand\subsection[1]{\refstepcounter{subsection}}%
   \RenewEnviron{theorem}{\refstepcounter{theorem}}%
  }%
  {}

\usepackage{xargs}
\usepackage[pdftex,dvipsnames]{xcolor}
\usepackage[colorinlistoftodos,prependcaption,textsize=tiny]{todonotes}
\newcommandx{\unsure}[2][1=]{\todo[linecolor=red,backgroundcolor=red!25,bordercolor=red,#1]{#2}}
\newcommandx{\change}[2][1=]{\todo[linecolor=blue,backgroundcolor=blue!25,bordercolor=blue,#1]{#2}}
\newcommandx{\info}[2][1=]{\todo[linecolor=OliveGreen,backgroundcolor=OliveGreen!25,bordercolor=OliveGreen,#1]{#2}}
\newcommandx{\improvement}[2][1=]{\todo[linecolor=Plum,backgroundcolor=Plum!25,bordercolor=Plum,#1]{#2}}
\newcommandx{\thiswillnotshow}[2][1=]{\todo[disable,#1]{#2}}

\usepackage{epstopdf,epsfig}  
\DeclareMathOperator*{\argmax}{argmax}
\DeclareMathOperator*{\E}{\mathbb{E}}

\newcommand{\interval}[1]{\square #1}
\newcommand{\imin}[1]{\underline{#1}}
\newcommand{\imax}[1]{\overline{#1}}
\newcommand{\eqdef}{\buildrel \text{def}\over =}
\newtheorem{assumption}{Assumption}
\newtheorem{lemma}{Lemma}
\newtheorem{remark}{Remark}
\newtheorem{theorem}{Theorem}
\newtheorem{property}{Property}

\title{Approximate Robust Control of Uncertain Dynamical Systems}

\author{
  Edouard~Leurent\\
  INRIA Lille, Renault\\
  \texttt{edouard.leurent@inria.fr} \\
  \And
  Yann~Blanco \\
   Renault \\
   \texttt{yann.blanco@renault.com} \\
   \AND
   Denis~Efimov \\
   Non-A team, INRIA Lille \\
   \texttt{denis.efimov@inria.fr} \\
   \And
   Odalric-Ambrym~Maillard \\
   SequeL team, INRIA Lille \\
   \texttt{odalric.maillard@inria.fr} \\
}

\begin{document}

%

\maketitle

\begin{abstract}
This work studies the design of safe control policies for large-scale non-linear systems operating in uncertain environments. In such a case, the robust control framework is a principled approach to safety that aims to maximize the worst-case performance of a system. However, the resulting optimization problem is generally intractable for non-linear systems with continuous states. To overcome this issue, we introduce two tractable methods that are based either on sampling or on a conservative approximation of the robust objective. The proposed approaches are applied to the problem of autonomous driving.
\end{abstract}

\section{Introduction}

\vspace{-3mm}
\noindent
Reinforcement Learning is a general framework that allows the optimal control of a Markov Decision Process $(\mathcal{S}, \mathcal{A}, T, r)$ with state space $\mathcal{S}$, action space $\mathcal{A}$, reward function $r\in [0, 1]^{\mathcal{S}\times\mathcal{A}}$ and unknown transition dynamics $T(s' | s, a) \in \mathcal{M}(\mathcal{S})^{\mathcal{S}\times\mathcal{A}}$ by searching for the policy $\pi \in \mathcal{M}(\mathcal{A})^\mathcal{S}$ with maximal expected value $v_\pi^T$ of the total discounted reward $R_\pi^T$:

\vspace{-5mm}
\begin{equation}
R_\pi^T(s) \eqdef \sum_{t=0}^\infty \gamma^t r(s_t, a_t), \qquad v_\pi^T (s) \eqdef \E({R_\pi^T (s)}),
\end{equation}

\vspace{-3mm}
\noindent
where $s_0=s$ , $a_t\sim\pi(s_t)$, $s_{t+1}\sim T(s_{t+1} | s_t, a_t)$, $\gamma\in[0, 1)$ is the discount factor and $\mathcal{M}(X)$ denotes the set of probability measures over $X$.

Unfortunately, its application to real-world tasks has so far been limited by its considerable need for experiences. It is generally recognized \citep{Sutton1990,Atkeson1997} that the most sample-efficient approach is the family of model-based methods which learn a nominal model $\hat{T}$ of the environment dynamics that is leveraged for policy search:

\vspace{-5mm}
\begin{equation}
\label{nominal-control-eq}
\max_\pi v_\pi^{\hat{T}}
\end{equation}

\vspace{-3mm}
\noindent
One drawback of such methods is that they suffer from model bias; that is, they ignore the error between the learned dynamics $\hat{T}$ and the real environment $T$. It has been shown that model bias can dramatically degrade the policy performances \citep{Schneider1997}.

\vspace{-1mm}
Model errors can instead be explicitly considered and expressed through an \textit{ambiguity set} of all possible dynamics models
. Such a set can be constructed from a history of observations by computing the confidence regions associated with the system identification process \citep{Iyengar2005, Nilim2005, Dean2017, Maillard2017}. In this work, we will consider ambiguity sets of parametrized deterministic dynamical systems $s' = T_\theta(s, a)$ whose unknown parameters $\theta$ lie in a compact set $\Theta$ of $\mathbb{R}^p$.

\vspace{-1mm}
In the optimal control framework, model uncertainty is handled by maximizing the \textit{expected} performances with respect to unknown dynamics. In stark contrast, in real-world applications where failures may turn out very costly, the decision maker often prefers to minimize the risk of the policy, which can be defined with several metrics characterizing the distribution of the policy outcome \citep{Garcia2015}.

\vspace{-1mm}
The robust control framework is a popular setting in which the risk of a policy is defined as the worst possible outcome realization among the ambiguity set, to guarantee a lower-bound performance of the robust policy when executed on the true model:  

\vspace{-5mm}
\begin{equation}
\label{eq:robust-control}
\max_\pi \min_T v^T_\pi
\end{equation}

\vspace{-3mm}
\noindent
Robust optimization has been studied in the context of finite Markov Decision Processes (MDP) with uncertain parameters by \citet{Iyengar2005}, \citet{Nilim2005} and \citet{Wiesemann2013}. They show that the main results of Dynamic Programming can be extended to their robust counterparts only when the dynamics ambiguity set verifies certain rectangularity properties. In the control theory community, the robust control problem is mainly restricted to the context of linear dynamical systems with bounded uncertainty in the time or frequency domain, where the objective is to guarantee stability \citep[e.g. $\mathcal{H}_\infty$-optimal control, see][]{Basar1996} or performance \citep[e.g. LQ optimal control theory, see][]{Petersen2014}. The existing nonlinear robust control approaches such as sliding mode control \citep{Li2018}, feedback linearization, backstepping, passivation and input-to-states stabilization \citep{Khalil2015} are usually based on canonical representations of regulated dynamics and admit constructive numeric realizations for systems of rather low dimensions.

\vspace{-1mm}
There have been few attempts of robust control of large-scale systems with both continuous states and non-linear dynamics, which is the focus of this paper. Our contribution is twofold. In section \ref{sampling}, we first consider a simpler case where the ambiguity set $\Theta$ and action space $\mathcal{A}$ are both finite and introduce a sampling-based planner that approximately maximizes the robust objective \eqref{eq:robust-control}. In section \ref{intervals}, we move to continuous ambiguity sets and form a conservative relaxation of the robust policy evaluation problem using interval predictors. In section \ref{experiments}, we illustrate the benefits of both techniques (for discrete, versus continuous $\Theta$) on a problem of tactical decision-making for autonomous driving.

\vspace{-4mm}
\section{Sampling-based planning}
\label{sampling}

\vspace{-2mm}
If the true dynamics model $T_\theta$ were known and the action-space $\mathcal{A}$ finite, sampling-based algorithms could be used to perform approximate optimal planning. In order to generalize to the robust setting, we need to make the following assumption about the structure of the ambiguity set:

\begin{assumption}[Structure]
The ambiguity set $\Theta$ and the action space $\mathcal{A}$ are discrete and finite:
\begin{equation}
\mathcal{A} = \{a_k\}_{k\in[1, K]} \quad \text{and} \quad \Theta = \{\theta_m\}_{m\in[1, M]}
\end{equation}
\end{assumption}

\vspace{-3mm}
\noindent
We slightly abuse notation and denote $T_m = T_{\theta_m}$.

\vspace{-1mm}
Such a structure of the ambiguity set typically stems directly from expert knowledge of the problem at hand. In general, it is nonrectangular, which implies that the Robust Bellman Equation does not hold \citep{Wiesemann2013}. This prevents us from building on planners that implicitly use this property and generate trajectories step-by-step by picking promising successor states, such as MCTS \citep{Coulom2006} or UCT \citep{Kocsis2006}.
Instead, we turn to algorithms that perform optimistic sampling of entire sequences of actions and work directly at the leaves of the expanded tree \citep[see, e.g.][]{Bubeck2010}. More precisely, we build on the work of \citet{Hren2008} on optimistic planning for deterministic dynamics, which we extend to the robust setting.

\vspace{-1mm}
We use similar notations and consider the infinite look-ahead tree $\mathcal{T}$ composed of all reachable states. Each node corresponds to a joint state $\{s_{m,t}\}_{m\in[1, M]}$ associated with the different dynamics $T_m$. The root starts at the current state, and all nodes have $K$ children, each corresponding to an action $a_k\in\mathcal{A}$ and associated with the successor joint state $\{s_{m,t+1}=T_m(s_{m,t}, a_k)\}_{m\in[1, M]}$. We use the standard notations over alphabets to refer to nodes in $\mathcal{T}$ as action sequences. Thus, a finite word $i \in \mathcal{A}^*$ of length $d$ represents the node obtained following the action sequence $(i_0, \cdots, i_d)$ from the root. Sequences $i\in\mathcal{A}^*$ and $j\in\mathcal{A}^*$ can be concatenated as $ij\in\mathcal{A}^*$, the set of suffixes of $i$ is $i\mathcal{A}^\infty = \{j\in\mathcal{A}^\infty: \exists h\in\mathcal{A}^\infty$ such that $j=ih\}$, and the empty sequence is $\emptyset$.

\vspace{-1mm}
The sample complexity is expressed in terms of number $n$ of expanded nodes. It is related to the number of calls to dynamics models: when a node $i$ is expanded, all successor states are computed for all $K$ actions and $M$ dynamics. At an iteration $n$, we denote $\mathcal{T}_n$ the tree of already expanded nodes, and $\mathcal{L}_n$ the set of its leaves.

\vspace{-2mm}
\begin{paragraph}{Definition}
Fix a dynamics model $m\in[1, M]$. \citet{Hren2008} define for any node $i\in \mathcal{T}$ of depth $d$ the optimal value $v_i^m$, its lower bound u-value $u_i^m$ and upper-bound b-value $b_i^m$. These variables depend on the dynamics $m$ and will therefore be referred to with a superscript $m$ notation.

We extend these dynamics-dependent variables to the robust setting, using superscript $r$ in notations. 
\vspace{-1mm}
\begin{itemize}
\item The robust value $v^r_i$ of a path $i \in \mathcal{A}^*$ as the restriction of \eqref{eq:robust-control} to policies that start with the action sequence~$i$:

\vspace{-8mm}
\begin{equation}
v^r_i \eqdef \max_{\pi \in i\mathcal{A^\infty}} \min_{m\in[1, M]} R^{T_m}_\pi
\end{equation}

\vspace{-3mm}\noindent
By definition, the robust value of \eqref{eq:robust-control} is recovered at the root $v^r_\emptyset = v^r$.

\vspace{-1mm}
Moreover, for $i \in \mathcal{T}_n \setminus \mathcal{L}_n$ we have 

\vspace{-3mm}
\begin{equation}
\label{eq:max_vr}
v_i^r = \max_{\pi\in i\mathcal{A}^\infty} \min_{m \in [1, M]} R^{T_m}_\pi = \max_{a\in\mathcal{A}}\max_{\pi\in ia\mathcal{A}^\infty} \min_{m \in [1, M]} R^{T_m}_\pi = \max_{a\in\mathcal{A}}v_{ia}^r
\end{equation}

\vspace{-1mm}
\item The robust u-value $u_i^r$ of a leaf node $i \in \mathcal{L}_n$ is the worst-case discounted sum of rewards $r_t=r(s_{m,t},i_t)$ from the root to $i$. It is then backed-up to the rest of the tree:

\vspace{-4mm}
\begin{equation}
\label{eq:ur}
u_i^r(n) \eqdef
\begin{cases}
\min_{m\in[1, M]} \sum_{t=0}^{d-1} \gamma^t r_t &\text{if } i \in \mathcal{L}_n \text{ ;}\\
\max_{a\in\mathcal{A}} u_{ia}^r(n) & \text{if } i \in \mathcal{T}_n \setminus \mathcal{L}_n
\end{cases}
\end{equation}

\vspace{-1mm}
\item Likewise, the robust b-value $b_i^r$ is defined at leaf nodes and backed-up to the rest of the tree:

\vspace{-4mm}
\begin{equation}
\label{eq:br}
b_i^r(n) \eqdef
\begin{cases}
u_i^r(n) + \frac{\gamma^d}{1-\gamma} &\text{if } i \in \mathcal{L}_n \text{ ;}\\
\max_{a\in\mathcal{A}} b_{ia}^r(n) & \text{if } i \in \mathcal{T}_n \setminus \mathcal{L}_n 
\end{cases}
\end{equation}

\vspace{-2.5mm}\noindent
An illustration of the computation of the robust b-values is presented in Figure \ref{fig:drop}.

\begin{figure}
\centering
\begin{minipage}{.49\textwidth}
\centering
\includegraphics[width=\linewidth]{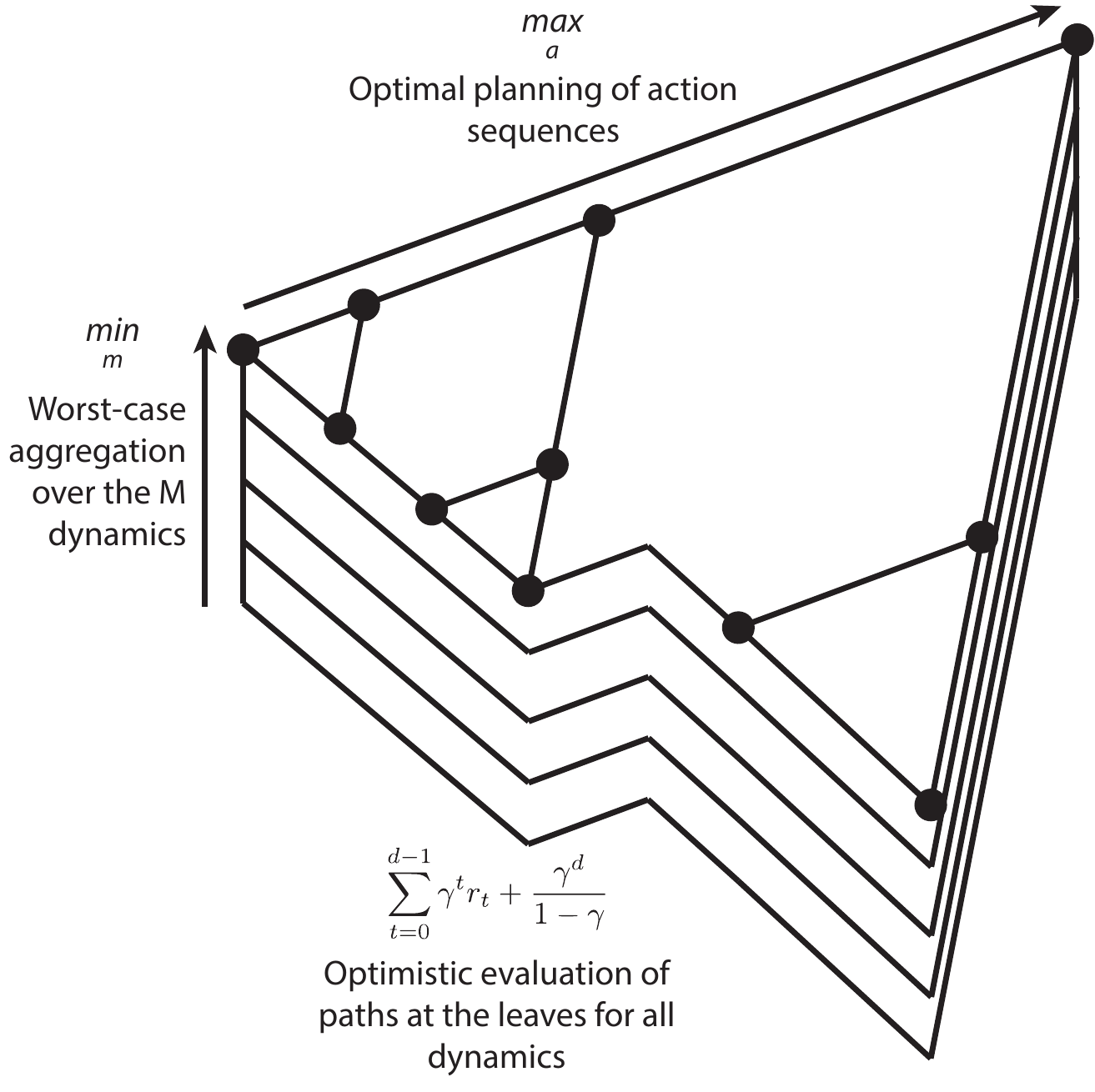}
\captionof{figure}{The computation of robust b-values in Algorithm \ref{algo:drop}. The simulation of trajectories for every dynamics model $T_m$ is represented as stacked versions of the expanded tree $\mathcal{T}_n$.}
\label{fig:drop}
\end{minipage}%
\hfill
\begin{minipage}{.49\textwidth}
\centering
\includegraphics[width=\textwidth]{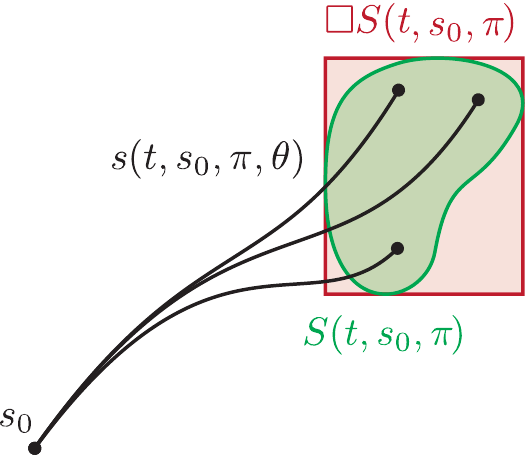}
\captionof{figure}{A few trajectories are sampled from an initial state $s_0$ following a policy $\pi$ with various dynamics parameters $\theta_m$ (in black). The union of reachability sets is shown in green, and its interval hull in red.}
\label{fig:interval-hull}
\end{minipage}

\vspace{-5mm}
\end{figure}

\begin{figure}[tp]
\centering

\end{figure}

\end{itemize}
\end{paragraph}

\vspace{-1mm}
\begin{remark}[On the ordering of min and max]
In the definition of $u_{i}^{r}(n)$ it is essential that the minimum among the models is only taken at the end of trajectories, in the same way as for the robust objective \eqref{eq:robust-control} in which the worst-case dynamics is only determined after the policy has been fully specified. Assume that $u_{i}^{r}(n)$ is instead naively defined as:

\vspace{-5mm}
\[
u_{i}^{r}(n)=\min_{m\in[1,M]}u_{i}^{m}(n),
\]

\vspace{-3mm}\noindent
This would not recover the robust policy, as we show in Figure \ref{fig:min-max-order} with a simple counter-example.
\begin{figure}[htp]
    \centering
    \includegraphics[width=\textwidth]{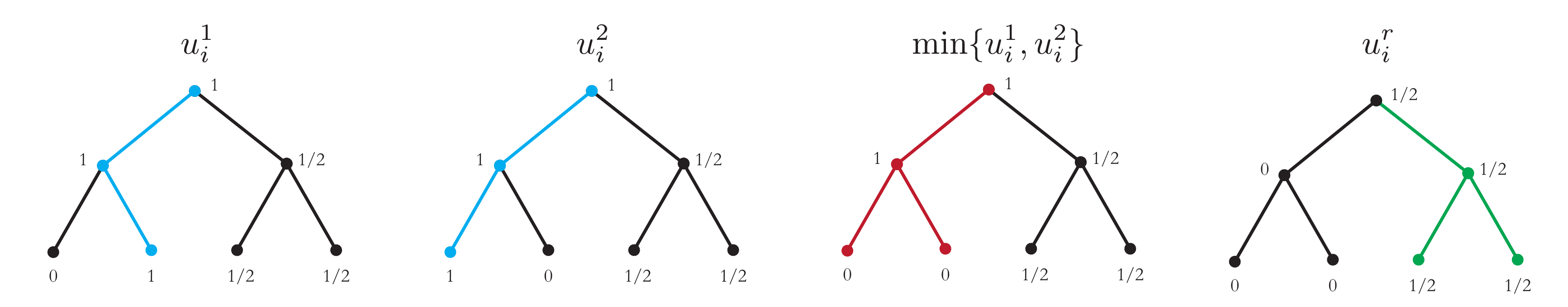}
    \caption{From left to right: two simple models and corresponding u-values with optimal sequences in blue; the naive version of the robust values returns sub-optimal paths in red; our robust u-value properly recovers the robust policy in green.}
    \label{fig:min-max-order}
    
    \vspace{-3mm}
\end{figure}
\end{remark}

\vspace{-2mm}
From these definitions we introduce Algorithm \ref{algo:drop}, and analyse its sample-efficiency in Theorem \ref{theorem:drop-regret}.

\begin{algorithm}[tp]
\DontPrintSemicolon
Initialize $\mathcal{T}$ to a root and expand it. Set $n=1$.\;
\While{Numerical resource available}{
Compute the robust u-values $u^r_i(n)$ and robust b-values $b^r_i(n)$.\;
Expand $\argmax_{i\in \mathcal{L}_n} b^r_i(n)$.\;
n = n + 1\;
}
\Return $\argmax_{a\in \mathcal{A}} u^r_a(n)$
\caption{Deterministic Robust Optimistic Planning}
\label{algo:drop}
\end{algorithm}

\begin{lemma}[Robust values ordering]
The robust values, u-values and b-values exhibit similar properties as the optimal values, u-values and b-values, that is: for all $0 < t < n$ and $i\in\mathcal{T}_n$,
\begin{equation}
u^r_i(t) \leq u^r_i(n) \leq v^r_i \leq b^r_i(n) \leq b^r_i(t)
\end{equation}
\label{lemma:uvb}
\end{lemma}

\vspace{-5mm}
\begin{proof}
This result stems directly from the definitions, see more details in Appendix \ref{appendix:proof-lemma}.
\end{proof}

\vspace{-1mm}
The simple regret of the action $a$ returned by Algorithm \ref{algo:drop} after $n$ rounds is defined as:
\begin{equation}
\mathcal{R}_n = v^r - v_a^r
\end{equation}
We will say that $\mathcal{R}_n=O(\varepsilon)$ for some $\varepsilon>0$ if there exist $\rho>0$ and $n_0>0$ such that $\mathcal{R}_n\leq\rho\varepsilon$ for all $n\geq n_0$.
A node $i\in\mathcal{T}$ is said to be $\epsilon$-optimal, in a robust sense, if and only if $v_i^r \geq v^r - \epsilon$ for some $\epsilon > 0$. The proportion of $\epsilon$-optimal nodes at depth $d$ is then defined as $p_d(\epsilon) = |i \in \mathcal{A}^d$ s.t $i$ is $\epsilon$-optimal$|/K^d$. Further we will assume that for the graph $\mathcal{T}$ the following hypothesis is satisfied:
\begin{assumption}[Proportion of near-optimal nodes]
\label{assumpt:beta}
There exist $\beta\in[0, \frac{\log K}{\log 1/\gamma}]$, $c > 0$ and $d_0 > 0$ such that $p_d(\epsilon)\leq c\epsilon^\beta$ for all $\epsilon > 0$ and $d\geq d_0$.
\end{assumption}

\begin{theorem}[Regret bound]
\label{theorem:drop-regret}
Let $\kappa = K\gamma^\beta \in [1, K]$. Then the simple regret of Algorithm \ref{algo:drop} is:

\vspace{-5mm}
\begin{equation}
\text{If } \kappa>1,\qquad 
\mathcal{R}_n = O\left(n^{-\frac{\log 1/\gamma}{\log \kappa}}\right)
\end{equation}

\vspace{-4mm}
\begin{equation}
\text{If }\kappa=1,\qquad
\mathcal{R}_n = O\left(\gamma^{\frac{(1-\gamma)^\beta}{c}n}\right)
\end{equation}
\end{theorem}

\vspace{-5mm}
\begin{proof}
We use the properties shown in Lemma \ref{lemma:uvb} and derive a robust counterpart of the proof of \citet{Hren2008}, which we only modify slightly. See more details in Appendix \ref{appendix:proof-thm}
\end{proof}

\vspace{-3mm}
\section{Interval predictors}
\label{intervals}

\vspace{-2mm}
In this section, we assume that the ambiguity set $\Theta$ is continuous and bounded.

\vspace{-1mm}
In the robust objective \eqref{eq:robust-control}, the $\min$ operator only requires us to describe the set of states that can be reached with non-zero probability.

\vspace{-2mm}
\begin{paragraph}{Definition}
The \textbf{reachability set} $S$ at time $t$ is the set of all states that can be reached by starting from initial state $s_0 \in \mathcal{S}$ and following a policy $\pi \in \mathcal{A}^\mathcal{S}$ along the transition dynamics $T_\theta \in \mathcal{S}^{\mathcal{S}\times\mathcal{A}}$.

\vspace{-5mm}
\begin{equation}
S(t, s_0, \pi) \eqdef \{s_t: \exists \theta \in \Theta \text{ s.t. } s_{k+1} = T_\theta(s_{k}, a_k), a_k = \pi(s_{k}), k=0, \cdots, t-1\}
\end{equation}
\end{paragraph}

\vspace{-5mm}
This set can still have a complex shape. We approximate it by an overset easier to represent and manipulate: its interval hull.

\vspace{-2mm}
\begin{paragraph}{Definition}
The \textbf{interval hull} of $S$, denoted $\interval{S}=[\imin{s}, \imax{s}]$ is the smallest interval containing it: 

\vspace{-5mm}
\begin{equation}
\label{eq:max-interval}
\underline{s}(t, s_0, \pi) \eqdef \min S(t, s_0, \pi) \qquad \overline{s}(t, s_0, \pi) \eqdef \max S(t, s_0, \pi)
\end{equation}

\vspace{-3mm}\noindent
The max and min operators are applied element-wise. This set is illustrated in Figure \ref{fig:interval-hull}.

\end{paragraph}

State intervals $\interval{S}$ have been used to describe the evolution of uncertain systems and derive feedback laws that achieve closed-loop stability in the presence of bounded disturbances \citep{Stinga2012, Efimov2016, Dinh2017}.

\vspace{-1mm}
The main techniques of interval simulation have been listed and described in a survey by \citet{Puig2005}, in which they are sorted into two categories. Region-based methods use the estimate of $\interval{S}$ at previous timestep $t-1$  to bootstrap the current estimate at time $t$. They are based on application of the theory of positive systems, which are frequently computationally efficient. However, the positive inclusion dynamics of a system may lead to overestimations of the true $\interval{S}$ and even unstable behaviour. Trajectory-based methods estimate $\interval{S}$ by taking the $\max$ and $\min$ in \eqref{eq:max-interval} over sampled trajectories for $\theta \in \Theta$. These methods produce subset estimates of the true $\interval{S}$, do not suffer from the wrapping effect, but are often more computationally costly.

\vspace{-1mm}
In this work, we leverage them to derive a proxy for the robust objective \eqref{eq:robust-control}.

\vspace{-3mm}
\begin{paragraph}{Definition}
Let us denote the robust objective of equation \eqref{eq:robust-control} as $v^r(\pi) \eqdef \min_{\theta\in\Theta} v^{T_\theta}_\pi$.

\vspace{-1mm}
We define the \textbf{surrogate objective} $\hat{v^r}$ on a finite horizon $H > 0$ as: 

\vspace{-5mm}
\begin{equation}
\hat{v^r}(\pi) \eqdef \sum_{t=0}^H \gamma^t \min_{s\in \square S(t, s_0, \pi)}  r(s, \pi(s))
\end{equation}
\end{paragraph}

\vspace{-1mm}
\begin{algorithm}[tp]
  \SetAlgoLined\DontPrintSemicolon
  \SetKwFunction{algo}{robust\_control}\SetKwFunction{proc}{evaluate}
  \SetKwProg{myalg}{Algorithm}{}{}
  \myalg{\algo{$s_0$}}{
  Initialize a set $\Pi$ of policies\;
  \While{resources available}
  {
  \proc{} each policy $\pi\in\Pi$ at current state $s_0$\; 
  Update $\Pi$ by policy search\;  
  }
  \KwRet $\argmax_{\pi\in\Pi} \hat{v^r}(\pi)$ \;}{}
  \setcounter{AlgoLine}{0}
  \SetKwProg{myproc}{Procedure}{}{}
  \myproc{\proc{$\pi$, $s_0$}}{
  Compute the state interval $\interval{S}(t, s_0, \pi)$ on a horizon $t\in[0, H]$\;
  Minimize $r$ over the intervals $\interval{S}(t, s_0, \pi)$ for all $t\in[0, H]$\;
  \KwRet $\hat{v^r}(\pi)$\;}
\caption{Interval-based Robust Control}
\label{algo:irc}

\vspace{-1mm}
\end{algorithm}

\vspace{-2mm}
\begin{property}[Lower bound]
\label{prop:lower-bound}
The surrogate objective $\hat{v^r}$ is a lower bound of the true objective $v^r$:

\vspace{-5mm}
\begin{equation}
\forall \pi, \hat{v^r}(\pi) \leq v^r(\pi)
\end{equation}
\end{property}

\vspace{-5mm}
\begin{proof}
By bounding the collected rewards by their minimum over $\interval{S}(t)$. See Appendix \ref{appendix:proof-prop} 
\end{proof}

\vspace{-2mm}
The robust objective error $v^r - \hat{v^r}$ stems from two terms: the interval approximation of the reachable set and the loss of time-dependency between the states within a single trajectory. If both these approximations are tight enough, maximizing the lower bound $\hat{v^r}$ will increase the true objective $v^r$, which is the idea behind Algorithm \ref{algo:irc}. It is classically structured as an alternation of a Policy Evaluation step , during which the surrogate objective $\hat{v^r}(\pi)$ is evaluated for a set of policies $\Pi$, and a Policy Search step which aims to steer the set of policies $\Pi$ towards regions where the surrogate objective is maximal. The main Policy Search algorithms are listed in a survey by \citet{Deisenroth2011b}. In this case, derivative-free methods such as evolutionary strategies (e.g. CMAES)  would be more appropriate than policy gradient methods, since $\hat{v^r}$ cannot be easily differentiated. Planning algorithms can also be used to exploit the dynamics and structure of the surrogate objective.

\vspace{-3mm}
\section{Experiments}
\label{experiments}

\vspace{-3mm}
Most autonomous driving architectures perform sequentially the prediction of other drivers’ trajectories and the planning of a collision-free path for the ego-vehicle. As a consequence, they fail to account for interactions between the traffic participants and the ego-vehicle, leading to overly conservative decisions and a lack of negotiation abilities \citep{Trautman2010}.
In this work, we perform both tasks \emph{jointly} to anticipate the effect of our own decisions on the dynamics of the nearby traffic. But human decisions are not fully predictable and cannot be reduced to a single deterministic model. To avoid model bias, we provide a whole ambiguity set of reasonable closed-loop behavioural models for other vehicles, and plan robustly with respect to this ambiguity.

\vspace{-1mm}
We introduce a new environment for simulated highway driving and tactical decision-making.\footnote{Source code is available at \href{https://github.com/eleurent/highway-env}{https://github.com/eleurent/highway-env}}

\begin{figure}[tp]
\centering
\begin{subfigure}[t]{0.49\textwidth}
    \centering
    \includegraphics[width=\textwidth]{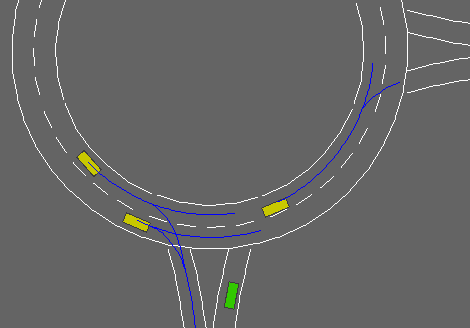}
    \caption{The possible trajectories (blue) for fixed behaviours and varying destinations}
    \label{highway-env-discrete}
\end{subfigure}
\begin{subfigure}[t]{0.49\textwidth}
    \centering
    \includegraphics[width=\textwidth]{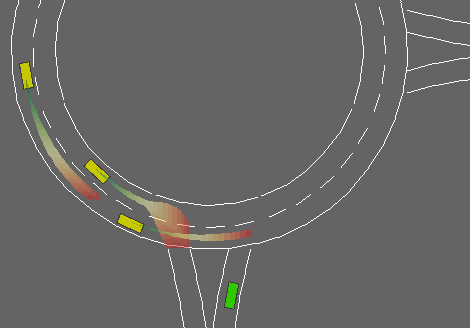}

    \caption{The possible trajectories (green-red gradient) for fixed destination and varying behaviours}
    \label{highway-env-interval}
    \end{subfigure}
    
\caption{The \href{https://github.com/eleurent/highway-env}{highway-env} environment. The ego-vehicle (green) is approaching a roundabout with flowing traffic (yellow).}
\label{highway-env}

\vspace{-5mm}
\end{figure}

\vspace{-1mm}
Vehicle motion is described by the Kinematic Bicycle Model \citep[see, e.g.][]{Polack2017}. They follow a lane keeping lateral behaviour, and a longitudinal behaviour inspired by the Intelligent Driver Model \citep{Treiber2000} which balances reaching a desired velocity and respecting a safe time gap. The lane-change decisions are determined by the MOBIL model \citep{Kesting2007}: they must increase the vehicles accelerations while satisfying safe braking decelerations. The behaviour parameters $\theta$ of each traffic participant are sampled uniformly from a set $\Theta$.

\vspace{-1mm}
The ego-vehicle can be controlled with a finite set of tactical decisions $\mathcal{A}$ = \{\texttt{no-op}, \texttt{right-lane}, \texttt{left-lane}, \texttt{faster}, \texttt{slower}\} implemented by low-lever controllers. It is rewarded for driving fast along a planned route while avoiding collisions. More information on the environment modelling is provided in the appendices.

\vspace{-1mm}
We carry out two experiments\footnote{Video and source code are available at \href{https://eleurent.github.io/robust-control/}{https://eleurent.github.io/robust-control/}}:  First, the behavioural parameters of traffic participants are fixed but their planned routes are unknown: we enumerate every direction they can take at their next intersection (see Figure \ref{highway-env-discrete}) and plan robustly with respect to this finite ambiguity set using Algorithm~\ref{algo:drop}. Second, we assume on the contrary that the agents' planned routes are known but not their behavioral parameters (see Figure \ref{highway-env-interval}). We plan robustly with respect to this continuous ambiguity set using Algorithm \ref{algo:irc}. Crucially, the state intervals prediction is conditioned on the planned policy $\pi$.

\vspace{-1mm}
In both experiments, we compare the performance of the robust planner to an oracle model that has perfect knowledge of the systems dynamics, and to a nominal planner that plans optimistically with respect to a dynamics model sampled uniformly from the ambiguity set. Statistics are collected from 100 episodes with random environment initialization. Results are presented in Table \ref{tab:experiments}.


\begin{table}[tp]
    \centering
    \caption{Performances of robust planners on two ambiguous environments.}
    \label{tab:experiments}
    \begin{tabular}{c c c c} \toprule
         Ambiguity set & Agent & Worst-case return & Mean return $\pm$ std  \\ \midrule
         True model & Oracle & $9.83$ & $10.84 \pm 0.16$ \\ \midrule
        \multirow{2}{*}{Discrete} & Nominal & $2.09$ & $8.85 \pm 3.53$ \\
         & Algorithm \ref{algo:drop} & \textbf{$8.99$} &  \textbf{$10.78 \pm 0.34$} \\ \midrule
        \multirow{2}{*}{Continuous} & Nominal & $1.99$ & $9.95 \pm 2.38$ \\
         & Algorithm \ref{algo:irc} & \textbf{$7.88$} & \textbf{$10.73 \pm 0.61$} \\ \bottomrule
    \end{tabular}

\vspace{-5mm}
\end{table}

\vspace{-3mm}
\section{Conclusion}

\vspace{-2mm}
This paper has presented two methods for approximately solving the robust control problem. In the simpler case of finite ambiguity set and action space, we use optimistic planning and provide an upper bound for the simple regret. A direct consequence is that we recover the robust policy as the computational budget increases. In the general case, we use interval prediction to efficiently solve a conservative approximation of the robust objective while providing a lower bound for the performance of a policy when applied to the unknown true model. However, this method is lossy and does not enjoy asymptotic consistency. Both algorithms are flexible, allowing to handle a variety of parametrized dynamical systems, and practical, with a focus on computational efficiency. The two methods are also orthogonal, which means they can be combined to deal with complex ambiguity sets that display both continuous and discrete features, such as disjoint unions of connected sets.

\vspace{-2mm}
\subsubsection*{Acknowledgments}

\vspace{-2mm}\noindent
This work has been supported by CPER Nord-Pas de Calais/FEDER DATA Advanced data science and technologies 2015-2020, the French Ministry of Higher Education and Research, INRIA, and the French Agence Nationale de la Recherche (ANR).

\small

\bibliographystyle{plainnat}
\bibliography{biblio}

\newpage

\begin{center}
\LARGE Supplementary material
\end{center}

\appendix

\section{Detailed proofs}

\subsection{Lemma \ref{lemma:uvb}}
\label{appendix:proof-lemma}
\begin{proof}

By definition, when starting with sequence $i$, the value $u_i^m(n)$ represents the minimum admissible reward, while $b_i^m(n)$ corresponds to the best admissible reward achievable with respect to the the possible continuations of $i$. Thus, for all $i\in\mathcal{A}^*$, $u_i^m(n)$ and $u_i^r(n)$ are non-decreasing functions of $n$ and $b_i^m(n)$ and $b_i^r(n)$ are a non-increasing functions of $n$, while $v_i^m$ and $v_i^r$ do not depend on $n$.

Moreover, since the reward function $r$ is assumed to have values in $[0, 1]$, the sum of discounted rewards from a node of depth $d$ is at most $\gamma^d + \gamma^{d+1}+\cdots = \frac{\gamma^d}{1-\gamma}$. As a consequence, for all $n \geq 0$ , $i\in\mathcal{L}_n$ of depth $d$, and any sequence of rewards $(r_t)_{t\in\mathbb{N}}$ obtained from following a path in $i\mathcal{A}^\infty$ with any dynamics $m \in [1, M]$:

\begin{equation*}
\sum_{t=0}^{d-1} \gamma^t r_t \leq \sum_{t=0}^{d-1} \gamma^t r_t + \sum_{t=d}^\infty \gamma^t r_t \leq \sum_{t=0}^{d-1} \gamma^t r_t + \frac{\gamma^d}{1-\gamma}
\end{equation*}
That is equivalent to:
\begin{equation*}
u^m_i(n) \leq \sum_{t=0}^\infty \gamma^t r_t \leq b^m_i(n) 
\end{equation*}
Hence,
\begin{equation}
\label{eq:min_m_values}
\min_{m \in [1, M]} u^m_i(n) \leq \min_{m \in [1, M]} \sum_{t=0}^\infty \gamma^t r_t \leq \min_{m \in [1, M]} b^m_i(n)
\end{equation}
And as the left-hand and right-hand sides of \eqref{eq:min_m_values} are independent of the particular path that was followed in $i\mathcal{A}^\infty$, it also holds for the robust path:

\begin{equation*}
\min_{m \in [1, M]} u^m_i(n) \leq \max_{\pi\in i\mathcal{A}^\infty} \min_{m \in [1, M]} \sum_{t=0}^\infty \gamma^t r_t \leq \min_{m \in [1, M]} b^m_i(n)
\end{equation*}
that is,
\begin{equation}
\label{eq:urvrbr}
u^r_i(n) \leq v^r_i  \leq b^r_i(n)
\end{equation}

Finally, \eqref{eq:urvrbr} is extended to the rest of $\mathcal{T}_n$ by recursive application of \eqref{eq:max_vr}, \eqref{eq:ur} and \eqref{eq:br}.
\end{proof}

\subsection{Theorem \ref{theorem:drop-regret}}
\label{appendix:proof-thm}

\begin{proof}
\citet{Hren2008} first show in Theorem 2 that the simple regret of their optimistic planner is bounded by $\frac{\gamma^{d_n}}{1 - \gamma}$ where $d_n$ is the depth of $\mathcal{T}_n$. This properties relies on the fact that the returned action belongs to the deepest explored branch, which we can show likewise by contradiction using Lemma \ref{lemma:uvb}. This yields directly that $a = i_0$ where $i$ is some node of maximal depth $d_n$ expanded at round $t\leq n$, which by Algorithm \ref{algo:drop} verifies $b_a^r(t) = b_i^r(t) = \max_{x\in\mathcal{A}} b_x^r(t)$ and:
\begin{equation}
\label{eq:Rndn}
v^r - v_a^r = v_{a^*}^r - v_a^r \leq b_{a^*}^r(t) - v_a^r \leq b_{a}^r(t) - u_a^r(t) = b_{i}^r(t) - u_i^r(t) = \frac{\gamma^{d_n}}{1-\gamma}
\end{equation}

Secondly, they bound the depth $d_n$ of $\mathcal{T}_n$ with respect to $n$. To that end, they show that the expanded nodes always belong to the sub-tree $\mathcal{T}_\infty$ of all the nodes of depth $d$ that are $\frac{\gamma^d}{1-\gamma}$-optimal. Indeed, if a node $i$ of depth $d$ is expanded at round $n$, then $b_i^r(n) \geq b_j^r(n)$ for all $j\in \mathcal{L}_n$ by Algorithm \ref{algo:drop}, thus the max-backups of \eqref{eq:br} up to the root yield $b^r_i(n) = b_\emptyset^r(n)$. Moreover, by Lemma \ref{lemma:uvb} we have that $b_\emptyset^r(n) \geq v_\emptyset^r = v^r$ and so $v_i^r \geq u_i^r(n) = b_i^r(n) - \frac{\gamma^d}{1-\gamma} \geq v^r - \frac{\gamma^d}{1-\gamma}$, thus $i \in \mathcal{T}_\infty$.

Then from Assumption \ref{assumpt:beta} and the definition of $\beta$ applied to nodes in $\mathcal{T}_\infty$, there exists $d_0$ and $c$ such that the number $n_d$ of nodes of depth $d \geq d_0$ in $\mathcal{T}_\infty$ is bounded by $c\left(\frac{\gamma^d}{1-\gamma}\right)^\beta K^d$. As a consequence, 
\begin{eqnarray*}
n &= \sum_{d=0}^{d_n} n_d = n_0 + \sum_{d=d_0+1}^{d_n} n_d \\
 &\leq n_0 + \sum_{d=d_0+1}^{d_n} c\left(\frac{\gamma^d}{1-\gamma}\right)^\beta K^d \\
 &= n_0 + c'\sum_{d={d_0+1}}^{d_n} \kappa^d
\end{eqnarray*}
 where $c'=\frac{c}{(1-\gamma)^\beta}$.

\begin{itemize}
    \item If $\kappa > 1$, then $n \leq n_0 + c'\kappa^{d_0+1}\frac{\kappa^{d_n-d_0}-1}{\kappa-1}$ and thus $d_n \geq d_0 + \log_\kappa \frac{(n-n_0)(\kappa - 1)}{c'\kappa^{d_0+1}}$.
We conclude from \eqref{eq:Rndn} that $\mathcal{R}_n \leq \frac{\gamma^{d_n}}{1-\gamma} = \frac{1}{1-\gamma} \left( \frac{(n-n_0)(\kappa - 1)}{c'\kappa^{d_0+1}} \right)^\frac{\log \gamma}{\log \kappa} = O\left(n^{-\frac{\log 1/\gamma}{\log \kappa}}\right)$.

\item If $\kappa = 1$, then $n \leq n_0 + c'(d_n-d_0)$, hence from \eqref{eq:Rndn} we have $\mathcal{R}_n = O\left(\gamma^{nc'}\right)$.
\end{itemize}
\end{proof}

\subsection{Property \ref{prop:lower-bound}}
\label{appendix:proof-prop}

\begin{proof}
For any $\theta \in \Theta$, $t \in [0, H]$ and any trajectory $(s_0, \cdots, s_t)$ sampled from $\pi$ and $T_\theta$,

\begin{equation*}
s_t \in S(t, s_0, \pi) \subset \square S(t, s_0, \pi)
\end{equation*}

Hence, 
\begin{equation*}
R^{T_\theta}_\pi = \sum_{t=0}^\infty \gamma^t r(s_t, a_t) \geq \sum_{t=0}^H \gamma^t r(s_t, a_t) \geq \sum_{t=0}^H \min_{s\in \square S(t, s_0, \pi)} \gamma^t r(s, \pi(s)) = \hat{v^r}(\pi)
\end{equation*}

And finally,
\begin{align*}
v^r(\pi) = \min_{\theta\in\Theta} {v_\pi^{T_\theta}} = \min_{\theta\in\Theta} \E(R^{T_\theta}_\pi) \geq \hat{v^r}(\pi)
\end{align*}
\end{proof}

\section{Environment dynamics}
	
\subsection{Kinematics}

The vehicles kinematics are represented by the Kinematic Bicycle Model:

\begin{align}
\dot{x} &= v\cos(\psi), \label{eq:kinematics-x}\\
\dot{y} &= v\sin(\psi), \label{eq:kinematics-y}\\
\dot{v} &= a, \label{eq:kinematics-v} \\
\dot{\psi} &= \frac{v}{l}tan(\beta), \label{eq:kinematics-psi}
\end{align}

where $(x, y)$ is the vehicle position, $v$ its forward velocity and $\psi$ its heading, $l$ is the vehicle half-length, $a$ is the acceleration command and $\beta$ is the slip angle at the center of gravity, used as a steering command.

Each vehicle $i$ is represented by its kinematics $X_i = [x_i, y_i, v_i, \psi_i]$. The joint state is represented by $s = \{X_1, \cdots, X_N\}$

\subsection{Longitudinal control}

The acceleration control is assumed to be linearly parametrized:

\begin{equation}
a = \theta_a^T \phi_a(s, i),
\label{eq:theta_a}
\end{equation}

where $\theta_a$ is an uncertain weight vector, and $\phi_a(s, i)$ is a feature vector that depends on the joint state $s$ and considered vehicle $i$.

It is composed of:
\begin{itemize}
\item a target velocity seeking term,
\item a braking term to adjust velocity w.r.t. the front vehicle ,
\item a braking term to respect a safe distance w.r.t. the front vehicle.
\end{itemize}

Denoting $f_i$ the front vehicle preceding vehicle $i$, $\phi_a$ is defined by

\begin{equation}
\label{eq:phi_a}
\phi_a(s, i) = \begin{bmatrix}
v_0 - v_i \\
n(v_{f_i}-v_i) \\
n(x_{f_i} - x_i - (d_0 + v_iT)) \\
\end{bmatrix}
\end{equation}

where $n$ is the negative part function $n(x) = \min(x, 0)$ and $v_0, d_0$ and $T$ respectively denote the speed limit, jam distance and time gap given by traffic rules.

We observe that this model exhibits similar qualitative behaviours to the IDM's.

\subsection{Lateral control}

A non-linear lane-keeping controller is implemented as follows: a lane $L$ with lateral position $y_L$ and heading $\psi_L$ is tracked by performing

\begin{enumerate}
\item Position control
\begin{equation}
v_{y_{cmd}} = K_{p_y}(y_L-y)
\end{equation}
\item Lateral velocity to heading conversion
\begin{equation}
\psi_{ref} = \psi_L+\sin^{-1}\left(\frac{v_{y_{cmd}}}{v}\right)
\end{equation}
\item Heading control
\begin{equation}
\psi_{cmd} = K_{p_\psi}(\psi_{ref}-\psi)
\end{equation}
\item Heading rate to steering angle conversion
\begin{equation}
\beta = \tan^{-1}(\frac{l}{v}\psi_{cmd})
\end{equation}
\end{enumerate}

Finally,

\begin{equation}
\beta = \tan^{-1}(\frac{l}{v}K_{p_\psi}(\psi_L+\sin^{-1}\left(K_{p_y}\frac{y_L-y}{v}\right)-\psi))
\end{equation}

This non-linear controller presented in subsection can be linearised around its equilibrium $(y, \psi) = (y_L, \psi_L)$.

\begin{align}
\frac{l}{v}\tan \beta &= K_{p_\psi}(\psi_L+\sin^{-1}\left(K_{p_y}\frac{y_L-y}{v}\right)-\psi) \\
 &\simeq \frac{l}{v}(K_{p_\psi}(\psi_L+\left(K_{p_y}\frac{y_L-y}{v}\right)-\psi)) \\
&= \theta_b^T\phi_b
\label{eq:theta_b}
\end{align}

with 

\begin{equation}
\theta_b = \begin{bmatrix} K_{p_\psi} & K_{p_y}K_{p_\psi}\end{bmatrix}^T
\end{equation}

and
\begin{equation}
\label{eq:phi_b}
\phi_b = \begin{bmatrix}
\psi_L-\psi\\
\frac{1}{v}(y_L-y)
\end{bmatrix}
\end{equation}

\subsection{Discrete behaviour}

The MOBIL model \citep{Kesting2007}, which stands for \textit{Minimizing Overall Braking Induced by Lane Changes}, is a discrete lateral decision model that formulates a criterion for lane changes in terms of safe braking decelerations and increased overall accelerations according to a longitudinal model.

It states that a lane change should be performed if and only if:
\begin{enumerate}
\item It does not impose an unsafe braking on the target lane following vehicle:
\begin{equation}
\dot{v}_\text{rear} \geq -b_{\text{safe}}
\end{equation}
\item It enables the vehicle and (with a politeness factor $p$) its following vehicles on both current and target lanes to increase their overall acceleration:
\begin{equation}
\Delta\dot{v} + p(\Delta\dot{v}_\text{rear, current} + \Delta\dot{v}_\text{rear, target}) \geq a_{min}
\end{equation}
\end{enumerate}

This model describes changes in the target lane $L$.

\section{Interval Predictor}

\label{interval-observer}

In this section, we design an interval predictor for our system.

\subsection{Notations}

For any real variable $z$, we denote an interval containing $z$ as $\interval{z} = [\imin{z}, \imax{z}]$, such that $\imin{z} \leq z \leq \imax{z}$. As elements of $\mathbb{R}^2$, they can be scaled and offset by scalars. This definition is extended element-wise to vector variables.

Then, we define several operators over intervals $\interval{a} = [\imin{a}, \imax{a}]$ and $\interval{b} = [\imin{b}, \imax{b}]$

\begin{itemize}
\item The product operator $\times$ 
\begin{align}
\interval{a} \times \interval{b} = [&p(\imin{a})p(\imin{b})-p(\imax{a})n(\imin{b})-n(\imin{a})p(\imax{b})+n(\imax{a})n(\imax{b}), \\
&p(\imax{a})p(\imax{b})-p(\imin{a})n(\imax{b})-n(\imax{a})p(\imin{b})+n(\imin{a})n(\imin{b})]
\end{align}

where $p(\cdot)$ and $n(\cdot)$ are the projections onto $\mathbb{R}^+$ and $\mathbb{R}^-$, respectively.

\item The difference operator $-$
\begin{equation}
\interval{a} - \interval{b} = [\imin{a} - \imax{b}, \imax{a} - \imin{b}]
\end{equation}

\item The cosine and sine operators
\begin{align}
\cos(\interval{z}) = [-&1 \text{ if } \imin{z} \leq \pi \leq \imax{z} \text{ else } \min(\cos(\imin{z}), \cos(\imax{z})),\\
&1 \text{ if } \imin{z} \leq 0 \leq \imax{z} \text{ else } \max(\cos(\imin{z}), \cos(\imax{z}))]
\end{align}

\begin{align}
\sin(\interval{z}) = [-&1 \text{ if } \imin{z} \leq -\frac{\pi}{2} \leq \imax{z} \text{ else } \min(\sin(\imin{z}), \sin(\imax{z})),\\
&1 \text{ if } \imin{z} \leq +\frac{\pi}{2} \leq \imax{z} \text{ else } \max(\sin(\imin{z}), \sin(\imax{z}))]
\end{align}

\item The inverse operator $/$ over a positive interval $\interval{z} > 0$
\begin{equation}
1 / \interval{z} = [1/\imax{z}, 1/\imin{z}]
\end{equation}

\item Any other function $f$ is assumed increasing on the interval $\interval{z}$ and is applied coefficient-wise
\begin{equation}
f(\interval{z}) = [f(\imin{z}), f(\imax{z})]
\end{equation}

\end{itemize}

We start with an initial estimate of the intervals over state variables $x_I, y_I, v_I$ and $\psi_I$. Typically, we use zero-width intervals centred on the current state observation. Likewise, any variable $z$ used in place of an interval corresponds to the zero-width interval $[z, z]$.

\subsection{Intervals for features}

We use \eqref{eq:phi_a} and \eqref{eq:phi_b} respectively to derive intervals for the features $\phi_a$ and $\phi_b$ from the intervals over the states.

We index the front vehicle intervals with the subscript $f$
\begin{equation}
\interval{\phi_a} = \begin{bmatrix}
v_0 - \interval{v} \\
n(\interval{v_{f}} - \interval{v}) \\
n(\interval{x_{f}} - \interval{x} - (d_0 + T\interval{v})) \\
\end{bmatrix}
\end{equation}

and

\begin{equation}
\interval{\phi_b} = \begin{bmatrix}
(1 ~ / ~ \interval{v}) \times (y_L - \interval{y})\\
\psi_L - \interval{\psi}
\end{bmatrix}
\end{equation}

\subsection{Intervals for controls}

The controls intervals are derived from \eqref{eq:theta_a} and \eqref{eq:theta_b}
\begin{align}
\interval{a} &= \interval{\theta_a}^T \times \interval{\phi_a} \\
\interval{\left(\frac{l}{v}\tan\beta\right)} &= \interval{\theta_b}^T \times \interval{\phi_b}
\end{align}

\subsection{Intervals for velocity and heading}
The velocity interval is derived from \eqref{eq:kinematics-v} and the heading interval from \eqref{eq:kinematics-psi}
\begin{align}
\interval{\dot{v}} &= \interval{a} \\
\interval{\dot{\psi}} &= \interval{\left(\frac{l}{v}\tan\beta\right)}
\end{align}

\subsection{Intervals for positions}
Likewise, the positions interval are derived from the kinematics \eqref{eq:kinematics-x} and \eqref{eq:kinematics-y}
\begin{align}
\interval{\dot{x}} = \interval{v} \times \cos(\interval{\psi})\\
\interval{\dot{y}} = \interval{v} \times \sin(\interval{\psi})
\end{align}


\end{document}